\documentstyle [11pt]{article}
\newcommand{\be}{\begin{equation}}
\newcommand{\ee}{\end{equation}}
\newcommand{\bea}{\begin{eqnarray}}
\newcommand{\eea}{\end{eqnarray}}

\newcommand{\ksi}{{\xi}}

\newcommand{\vs}[1]{\rule[- #1 mm]{0mm}{#1 mm}}

\begin{document}
\renewcommand{\thefootnote}{\fnsymbol{footnote}}

\newpage

\newpage
\setcounter{page}{0}
\rightline{LPTHE 94/35}
\rightline{ May 1994\vs{30}}
\begin{center}
{\LARGE {\bf{ Nambu mass hierarchies  in strings.}}} \\
\vspace{1cm}
{\large P.\ Bin\'etruy and E.\ Dudas} \\
{\em Laboratoire de Physique Th\'eorique et Hautes Energies\\
Universit\'e Paris-Sud, Bat. 211, F-91405 Orsay Cedex, France} \\[1cm]

\end{center}
\vs{30}

\centerline{ \bf{Abstract}}
\renewcommand{\thefootnote}{\arabic{footnote}}
\setcounter{footnote}{0}

\indent

We show that a recent proposal by Nambu to generate a hierarchy among Yukawa
couplings in the standard model may be easily implemented in superstring
models. In such models, two of the main ingredients of the Nambu
proposal find a natural explanation: minimising with respect to the
Yukawas amounts to a minimisation with respect to the underlying moduli
fields and a constraint on the Yukawas of the type of the Veltman
condition may be attributed to the relaxation process to a
phenomenologically viable string vacuum.

\vfill
\renewcommand{\thefootnote}{\arabic{footnote}}
\setcounter{footnote}{0}
\newpage
\section{Introduction.}
\vskip 1cm
The hierarchy observed between the different masses of the known
particles (5 to 6 orders of magnitude between the electron and
the top) which  translates into a hierarchy of
Yukawa couplings is certainly one of the most challenging issues at
stake in the standard model. Now that we know that the top quark  is
heavy and that its mass lies in the range of the electroweak unification scale,
the problem has been recently rephrased as to why the other quarks and
leptons are so much lighter than this scale. This does not seem however
to have led to any obvious solution yet and there is a definite need for
new ideas to tackle this issue.

In this situation, an interesting proposal has been put forward by Nambu
\cite{nambu}.  The idea consists in minimizing the vacuum energy density
with respect to the Yukawa couplings -- keeping all the other
parameters fixed, in particular the field vacuum expectation values
(vevs) and the other couplings  -- under the condition of vanishing
quartic and quadratic divergences
in the Higgs sector (the latter being the Veltman condition \cite{Veltman}).

In the toy model chosen by Nambu, the Veltman condition reads, in the
case of two Yukawa couplings $\lambda_1$ and $\lambda_2$:
\be
\lambda_1^2 + \lambda_2^2  = a^2, \label{eq:Nambu}
\ee
with $a$ a constant. This constrains both Yukawa couplings to the region
$[0,a]$.
Such a condition cancels the order $\Lambda^2$ contributions to the
vacuum energy, where $\Lambda$ is the cut-off. One is left in the scalar
potential with the $O(\ln \Lambda)$
contributions which read:
\be
V_1 = -A (\lambda_1^4 + \lambda_2^4 ) + O(\lambda_i^2 \ln \lambda_i^2)
\ee
Such a potential favors, in the case of $A>0$, large hierarchies of couplings:
$(\lambda_1,\lambda_2) = (a,0)$ or $(0,a)$ when one disregards the logarithmic
corrections. The generic effect of these logarithmic corrections is to
generate a non-zero value for the smaller Yukawa coupling. The key
ingredient for this mechanism to work is the sign of $A$: we will see in
what follows that it is naturally negative in the case of supersymmetric
models. Similarly, the sign of the logarithmic corrections is important
since it may otherwise destabilize the hierarchy \cite{Gherghetta}.

An interesting feature of this mechanism is that it easily extends to
the case of an arbitrary number of Yukawa couplings. In such a
situation, one of the Yukawa couplings is naturally much larger than all
the other ones. This of course would provide a very simple explanation
as to why the top quark is much heavier than the remaining quarks and
leptons of the standard model.

The Nambu proposal however raises several questions as to its range of
applicability. In the first place, is it valid to treat the Yukawa
couplings as dynamical variables? Also, what about the stability of the
Veltman condition under renormalisation? One might also wonder whether
it is licit to minimize with respect to the Yukawa couplings while
keeping the fields at their vacuum expectation values: this procedure
could hide some dangerous instabilities of the model.

Some of these potential problems might be cured by making the model
supersymmetric.
For example, the contribution of the scalar fields necessarily present
in supersymmetric models tends to give the right sign for the constant
$A$ introduced above and for the logarithmic radiative corrections.
Moreover, the question of the dynamical nature of the Yukawa
couplings finds a natural answer in superstring models. In
these models, it is well-known that Yukawa couplings have a non-trivial
dependence on the moduli fields which characterize the K\"ahler  (and
complex) structure of the compact manifold. In some instances
\cite{DFMS,DSWW}, such couplings appear through non-perturbative effects on
the string world-sheet and are typically of order $e^{-T}$ where $T$ is
the modulus whose vev determines the sigma model coupling. In the case
of $(2,2)$ vacua, the couplings are determined by holomorphic functions
of the moduli which also generate the K\"ahler metrics \cite{DKL}.
Also, it has been recently noted \cite{Ignatios}
that through wave function renormalisation due to massive modes, Yukawa
couplings receive moduli dependent corrections at one loop: these
corrections can be understood as threshold effects at the string scale
and they make the boundary conditions for Yukawa couplings moduli
dependent.

In this context, minimization with respect to the Yukawa couplings in the
low energy theory amounts to a minimization with respect to the moduli
fields of the underlying string theory. Two cases may arise:

i) when one reaches low energies, enough moduli remain undetermined
({\it i.e.} they correspond to a flat direction of the potential) so
that one can minimize with respect to {\it all} Yukawa couplings.

ii) at low energies, in particular after supersymmetry breaking has
taken place, there remains fewer moduli than Yukawa couplings. This
situation would typically yield new constraints on the Yukawas
which might prove to be useful by further constraining the Yukawa
parameter space.

In what follows, we will suppose that we are in the first case and will
minimize with respect to all the Yukawa couplings. We will first discuss
in Section 2 to what extent string models have properties which allow to
easily implement  the Nambu mechanism. In Section 3 we study in some
details a toy model which we find
representative of the general features of these models.
\vskip 1cm
\section{Nambu hierarchies in string models}
\vskip .5cm
In what follows we will use the formulation of the effective potential
\cite{eff}. To one loop order, this reads:
\be
V = V_{tree} + {1 \over 32 \pi^2} \left( \Lambda^2 Str \; M^2  - {1 \over 2}
Str \; M^4 \left[ \ln {\Lambda^2 \over M^2}+ {1\over 2} + O({M^2 \over
\Lambda^2}) \right]  \right), \label{eq:Veff}
\ee
where
\be
Str \; F(M^2) = \sum_J (-1)^{2J} (2J+1) F(M_J^2).
\ee
We will assume that supersymmetry is broken in a hidden sector. This
induces soft supersymmetry breaking terms of the order of the gravitino mass
$m_{3/2}$ in
the observable sector of quarks and leptons. In the general case, $Str
\; M^2$ is of the order of $m_{3/2}^2$. This yields dangerous terms in the
potential (\ref{eq:Veff}) of order $m_{3/2}^2 \Lambda^2$ which tend to
destabilize the hierarchy $m_{3/2}/M_{Pl}$. In view of this, we will
consider only the more realistic models where this leading contribution
vanishes and therefore where \cite{Ellwanger}:
\be
Str \; M^2 = O \left( {m_{3/2}^4 \over M_{Pl}^2} \right).
\label{eq:constraint}
\ee
This is in some sense a weaker form of the Veltman condition for the
models that we study. It yields quite generally a quadratic constraint
on the Yukawa couplings.

This last statement requires some explanation since it is known that
supersymmetry (possibly softly broken) gives mass relations which
give rise to cancellations in $Str M^2$. We will take for the sake of
illustration the following model, which is somewhat representative of
what is found in string models. The field content is a dilaton $S$, a
modulus $T$ and a set of $n$ chiral superfields $\Phi_i$. The couplings
are described by a K\"ahler potential:
\be
K = -\ln (S + \bar{S}) - 3 \ln (T+\bar{T} - |\Phi|^2)
\label{eq:Kahler}
\ee
where $|\Phi|^2 = \sum_i|\Phi_i|^2$, and a superpotential $W(\Phi_i)$
which is cubic in the matter fields $\Phi_i$.

Then global supersymmetry ensures automatic cancellation in $Str \; M^2$
of some of the terms depending on Yukawas but not of all of them.
Indeed, let us write the terms of dimension $2$ and $4$ as respectively
\bea
{\cal W} &=& {1 \over 9} {1 \over (S + \bar{S}) (T+\bar{T} - |\Phi|^2) }
\sum_{i,j=1}^N \left| {d^2W\over d\Phi_i d\Phi_j} \right|{}^2, \nonumber \\
\hat{V} &=& {1 \over 3} {1 \over (S + \bar{S}) (T+\bar{T} - |\Phi|^2)^2 }
\sum_{i=1}^N \left| {dW\over d\Phi_i} \right|{}^2.
\eea
They contribute to the scalar trace $Tr M_S^2$ and spinor trace $Tr M_F^2$
as follows \cite{BG}:
\bea
Tr M_S^2 &=& 2(5+{2N\over3}) {\hat{V}} + 2 {\cal W} + \cdots \nonumber \\
Tr M_F^2 &=& {14\over3} {\hat{V}} +  {\cal W} + \cdots.
\eea
We thus check that the terms of order $2$ cancel in the supertrace, as
required by global supersymmetry:
\be
Str\; M^2 = {2\over3}(1+2N) {\hat{V}} + \cdots
\ee
But there remain some terms of dimension $4$ which depend on the Yukawa
couplings: these terms are obviously of order $1/M_{Pl}^2$ and are
therefore not constrained by global supersymmetry.
For instance, in the case of the toy model that we will
consider below,
\be
W = {1\over3} \lambda_1 \Phi_1^3 + {1 \over 3} \lambda_2 \Phi_2^3
\ee
and $\hat{V} = {\hat{\lambda}}_1^2 |\hat{\Phi}_1|^4 + {\hat{\lambda}}_2^2
|\hat{\Phi}_2|^4$, where ${\hat{\lambda}}_1$ and ${\hat{\lambda}}_2$ are the
low-energy Yukawa parameters:
\be
{\hat{\lambda}}_i^2 = {1 \over 27} {1 \over <S+\bar{S}>} \lambda_i^2.
\label{eq:hat}
\ee
and the $\Phi_i$ fields have been rescaled to $\hat{\Phi}_i$ in order to
have normalized kinetic terms.

There are of course other terms in $Str M^2$ which come from a hidden
sector that we have not completely specified. Following our assumptions,
the contribution of these terms is at most of order
$m_{3/2}^4/M_{Pl}^2$. Any condition of the type of the Veltman condition
will therefore yield
\be
\lambda_1^2 |\Phi_1|^4 + \lambda_2^2 |\Phi_2|^4 \sim m_{3/2}^4
\label{eq:quad}
\ee
and assuming $<\Phi_1^2>$ and $<\Phi_2^2>$ of the order of $m_{3/2}^2$, we
obtain a constraint of the form (\ref{eq:Nambu}).

Assuming a cancellation of the leading terms in $Str M^2$, one is left
with the terms of order $M^4$ in the effective potential
(\ref{eq:Veff}) which we can write, up to a constant additive term:
\bea
V&=& -{1 \over 64 \pi^2} Str M^4\left[ \log {\Lambda^2 \over M^2} +
{1\over 2}+ O \left( {M^2 \over \Lambda^2}\right) \right] \nonumber \\
&=& -{1 \over 64 \pi^2} Str M^4 \log {\Lambda^2 \over m_{3/2}^2} +
{1 \over 64 \pi^2} Str M^4\left[ \log {M^2 \over m_{3/2}^2} -
{1\over 2}+ O\left({M^2 \over \Lambda^2}\right) \right] \nonumber \\
&=& -A Str M^4 + O(\lambda_i^n \log \lambda_i)
\label{eq:effM4}
\eea
where $A={1 \over 64\pi^2} \log (\Lambda^2/m_{3/2}^2)$ is by
construction positive ($\Lambda \gg m_{3/2}$). Because of the quadratic
constraint (\ref{eq:quad}), $Str M^4$ is quartic in the Yukawa couplings
and we thus obtain a potential which has a behaviour very similar to
(\ref{eq:Nambu}), as advocated by Nambu in his toy model.
\vskip 1cm
\section{A toy model}
\vskip .5cm
We consider in this section a simple toy model which, we believe,
includes the basic features of more realistic string models, in order to
see how the Nambu mechanism works in these models. It is beyond the
scope of this work to present realistic scenarios explaining for example
why the top quark is much heavier than the bottom quark and the tau
lepton or why the third family is heavier than the remaining two. But we
will see from the study of our model to what extent supersymmetry and
radiative corrections are important ingredients for generating a
hierarchy of the Nambu type.

 Our model contains two observable chiral superfields $\Phi_1$ and
$\Phi_2$ whose low energy supersymmetric couplings are described by  the
superpotential:
\be
W = {1 \over 3} \lambda_1 \Phi_1^3 + {1 \over 3} \lambda_2 \Phi_2^3
\label{eq:super}
\ee
and an unspecified hidden sector where we assume that supersymmetry is
broken spontaneously. We will avoid discussing the issue of
supersymmetry breaking although this might play an important role in the
realization of the Veltman condition.

In the observable sector under study here, supersymmetry breaking induces
soft supersymmetry breaking terms whose scale is determined by the
gravitino mass $m_{3/2}$.
We will include here scalar mass terms of the form\footnote{ In all
generality the soft supersymmetry breaking terms compatible with the
$Z_3 \times Z_3$ symmetry of the observable sector present in our model
include A-terms: $V' = A_1 \lambda_1 z_1^3 + A_1 \lambda_2 z_2^3 +
h.c.$. For the sake of simplicity and in order to be able to present
analytic computations, we will not include them in the discussion.}
\be
V_{soft} =  \mu_1^2|z_1|^2 + \mu_2^2 |z_2|^2
\ee
where $z_1,z_2$ are the scalar components of $\Phi_1,\Phi_2$
respectively. In the usual scenarios of supersymmetry breaking, the
scalar masses are universal in order to prevent flavour changing neutral
currents and they are of the order of the gravitino mass. We will
therefore assume
\be
\mu_1 = \mu_2 = \mu = O(m_{3/2}).
\ee
The scalar fields $z_1$, $z_2$  must have nonzero vevs $v_1$ and $v_2$.
Otherwise their two supersymmetric partners $\psi_1$ and $\psi_2$, of
respective masses $m_1 = 2 \lambda_1 v_1$ and $m_2= 2 \lambda_2 v_2$
would remain massless, a case of limited interest to us. At tree level,we will
therefore take $\mu^2 < 0$ and
\be
v_1 \equiv <\Phi_1> = {|\mu^2| \over 2 \lambda_1^2}, \;\;\;
v_2 \equiv <\Phi_2> = {|\mu^2| \over 2 \lambda_2^2}.
\label{eq:vev}
\ee

The procedure we adopt goes as follows:

{\bf i)} we compute the one-loop effective potential $V(z_1,z_2)$ and
the corresponding saddle point equations which determine $v_1$ and
$v_2$.

{\bf ii)} we then obtain the vacuum energy ${\cal E}_0 (\mu^2,
\lambda_1, \lambda_2, \Lambda)$ in this one-loop approximation.

{\bf iii)} we write a Veltman type condition for the cancellation of
quadratic divergences up to order $m_{3/2}^4 \Lambda^2/M_{Pl}^2$. As
discussed above and as we will return below, this might not be necessary
for the mechanism to work. The condition acts as a quadratic constraint
on the couplings $\lambda_1$ and $\lambda_2$.

{\bf iv)} we minimize ${\cal E}_0$ with respect to $\lambda_1$ and
$\lambda_2$ using the constraint. Consequently, the couplings
$\lambda_1$ and $\lambda_2$ are dynamically determined, and so are the
fermion masses $m_1$ and $m_2$.

Assuming the presence of a dilaton $S$ and at least one modulus $T$
coupled to the matter superfields $\Phi_i$ through a K\"ahler potential
of the type (\ref{eq:Kahler}), we find at the ground state a
supertrace of $M^2$ which reads, in the absence of soft supersymmetry
breaking terms,
\be
Str M^2 = {\alpha \over M_{Pl}^2} ( \lambda_1^2 |v_1|^4 + \lambda_2^2
|v_2|^4) + \beta R  + m_{3/2}^2 \sum_{n=1} \gamma_n \left( {m_{3/2}
\over M_{Pl}}\right)^{2n}.
\ee
The origin of the first term was discussed in the previous
section.\footnote{The couplings $\lambda_i$ used here are actually the
rescaled couplings $\hat{\lambda}_i$ of eq.(\ref{eq:hat}). Similarly,
the fields $\Phi_i$  whose vev is $v_i$ are in fact the rescaled fields
$\hat{\Phi}_i$ with normalized kinetic terms.}

The second term appears when one considers a background with nonzero
curvature $R$ \cite{R}. Finally, the last term corresponds to the
contribution of the hidden sector, proportional to $m_{3/2}^2$. As discussed
above, the remaining fields in the model are chosen in order that the
dangerous contribution of order $m_{3/2}^2$ in $Str M^2$ automatically
cancels.\footnote{Ferrara, Kounnas and Zwirner are presently
constructing such string models where the moduli sector of the theory is
chosen precisely in order to cancel this term \cite{SUSY94}.}

We will assume that the supertrace cancels to order $m_{3/2}^4$, {\em
i.e.}
\be
{\alpha \over M_{Pl}^2} (\lambda_1^2 |v_1|^4 + \lambda_2^2 |v_2|^4)
+ \beta R  + \gamma_1 {m_{3/2}^4 \over M_{Pl}^2} = 0
\label{eq:veltman}
\ee
where the possibly non-vanishing background curvature is automatically
of order $|v_i|^4 \sim m_{3/2}^4$ ($R = 4<V>/M_{Pl}^2$).
This yields a constraint of the type (\ref{eq:Nambu}) for the Yukawa
couplings. As will become clear in what follows, a strict cancellation
as in (\ref{eq:veltman}) is not necessary: it is sufficient for the
hierarchy to be generated that the left-hand side be bounded by a
quantity of order $m_{3/2}^4/M_{Pl}^2$.

The tree level scalar potential in the observable sector simply reads:
\be
V_0 = \mu^2(|\Phi_1|^2 + |\Phi_2|^2) + \lambda_1^2 |\Phi_1|^4 +
\lambda_2^2 |\Phi_2|^4
\ee
and the field vevs are given by (\ref{eq:vev}). It follows that the
vacuum energy ${\cal E}_0$, using the condition
(\ref{eq:veltman}), is proportional
to $m_{3/2}^4$. As any scale in a string model, $m_{3/2}$ is determined
dynamically through the vev of some scalar field, say the dilaton $S$.
It is an important aspect of the strategy that we adopt here to assume
that the minimisation procedure that leads to the determination of
$m_{3/2}$ is independent of the minimisation with respect to the Yukawa
couplings. We therefore consider $m_{3/2}$ as an independent constant.
There is then no dependence at tree level of ${\cal E}_0$ on the
couplings $\lambda_1$ and $\lambda_2$ and we need to go to the one-loop
level.

At one loop, we use the effective potential of eq.(\ref{eq:Veff})
where the cut-off $\Lambda$ is taken to be of order $M_{Pl}$. The
supertraces of $M^2$ and $M^4$ read:
\bea
Str M^2 &=& 4\mu^2 + {1 \over M_{Pl}^2} (\lambda_1^2 |\Phi_1|^4 +
\lambda_2^2 |\Phi_2|^4) + \gamma {m_{3/2}^4 \over M_{Pl}^2}
+ O\left( {m_{3/2}^6 \over M_{Pl}^4} \right), \nonumber \\
Str M^4 &=& 4\mu^4 + 16 \mu^2 (\lambda_1^2 |\Phi_1|^2 +
\lambda_2^2 |\Phi_2|^2) + 8  (\lambda_1^4 |\Phi_1|^4 +
\lambda_2^4 |\Phi_2|^4) \nonumber \\
        & & + O\left( {\Phi^6 \over M_{Pl}^2} \right)
\eea
Due to the presence of the quadratic divergence $\Lambda^2$, the terms
of order $1/M_{Pl}^2$ in $StrM^2$, and only them in leading order,
 will contribute to the
scalar potential at low energy. Neglecting for the moment terms
logarithmic in the Yukawa couplings, we obtain for the scalar potential
to order one-loop
\bea
V &=& C + \mu^2 \left(1- {\lambda_1^2 \over 4 \pi^2} \log {\Lambda^2
\over \mu_0^2}\right) |\Phi_1|^2
        + \mu^2 \left(1- {\lambda_2^2 \over 4 \pi^2} \log {\Lambda^2
\over \mu_0^2}\right) |\Phi_2|^2 \nonumber \\
  &+& \lambda_1^2 \left(\rho- {\lambda_1^2 \over 8\pi^2} \log {\Lambda^2
\over \mu_0^2}\right) |\Phi_1|^4
       + \lambda_2^2 \left(\rho- {\lambda_2^2 \over 8\pi^2} \log {\Lambda^2
\over \mu_0^2}\right) |\Phi_2|^4 \nonumber \\
  &+& O(\lambda_i^n \log \lambda_i)
\label{eq:potential}
\eea
where $C$ is a constant -- {\em i.e.} independent of $\lambda_i$ and
$\Phi_i$ -- $\mu_0$ is some low energy renormalisation scale
and $\rho = 1 + (\Lambda^2/M_{Pl}^2)/(32\pi^2)>1$. The
expression (\ref{eq:potential}) is symmetric under the exchange
$(\lambda_1,\Phi_1 \leftrightarrow \lambda_2,\Phi_2)$.

We will use the notations:
\be
x_i = {\lambda_i^2 \over 8 \pi^2} \log {\Lambda^2 \over \mu_0^2},\;\;
\ksi_i = {1 \over x_i} \left({1-2x_i \over \rho - x_i}\right)^2, \;\;
i=1,2. \label{eq:ksi}
\ee
In this case the vevs $v_1$ and $v_2$ are given by
\be
v_i^2 = - {\mu^2 \log(\Lambda^2 / \mu_0^2) \over 16 \pi^2 x_i} {1-2x_i \over
\rho - x_i} = - {\mu^2 \log(\Lambda^2 / \mu_0^2) \over 16 \pi^2} \left(
{\ksi_i \over x_i}\right)^{1/2}
\ee
and the vacuum energy dependence reads (neglecting an additive numerical
constant)
\be
{\cal E}_0 = -M^4 [\ksi_1(\rho-x_1) + \ksi_2(\rho-x_2)]
\label{eq:Ezero}
\ee
where $M^4 = \mu^4 \log(\Lambda^2 / \mu_0^2) / (32 \pi^2)$.

In these notations, the  condition (\ref{eq:constraint}) reads
\be
\ksi_1 + \ksi_2 = \tilde{a}.
\label{eq:contrainte}
\ee
This condition can be used to eliminate one variable and write the
effective potential as a function of $\ksi_1$ only ($x_1$ and $x_2$
being understood as implicit functions  of $\ksi_1$ through (\ref{eq:ksi}).

We will then show that ${\cal E}_0(\ksi_1)$ has a single extremum
corresponding to the symmetric solution $\ksi_1 = \ksi_2$. This extremum
being a maximum as we will see, the minimum value of ${\cal E}_0$ is
rejected to the boundary values for $\ksi_1$, {\em i.e.}
$\ksi_1=\tilde{a}$ ($\ksi_2 = 0$) or $\ksi_1 = 0$ ($\ksi_2 =
\tilde{a}$).

Starting with
\be
{d{\cal E}_0 \over d\ksi_1} = -M^4 \left( x_2 - x_1 - \ksi_1 {\partial x_1
\over \partial \ksi_1} - \ksi_2 {\partial x_2 \over \partial \ksi_1} \right)
\ee
and computing $\partial \ksi_1 / \partial x_i$ ($i=1,2$), one finds that
\be
{d{\cal E}_0 \over d\ksi_1} = - (x_2 - x_1) f(x_1,x_2)
\ee
where $f(x_1,x_2)$ is strictly positive for $x_i \leq 1/2$.
If $x_i \geq 1/2$ the only solution of the saddle point equations is
$v_1=v_2=0$ so the equations above no longer hold.

Having proved that the only extremum of the effective potential
corresponds to $x_1 = x_2$, {\em i.e.} to $\xi_1 = \xi_2 = \tilde{a}/2$,
the simplest way to proceed is to compare the energy of this solution
with the energy of another solution of the quadratic constraint
(\ref{eq:contrainte}), for
example $\xi_1 = 0,\xi_2 = \tilde{a}$ (or vice-versa).

In the symmetric case ($\xi_1 = \xi_2 = \tilde{a}/2$ hence
$x_1=x_2\equiv x_{sym}$), one finds from (\ref{eq:Ezero})
\be
{\cal E}_0|_{sym} = - M^4 \tilde{a} (\rho - x_{sym}).
\ee
In the asymmetric case ( say $\xi_1 = 0, \xi_2 = \tilde{a}$), one finds
\be
{\cal E}_0|_{asym} = -M^4 \tilde{a} (\rho - x_{asym})
\ee
where $x_{asym}$ is the corresponding value of $x_2$. It is easy to
prove that $x_{sym} > x_{asym}$. Hence ${\cal E}_0|_{asym}<{\cal
E}_0|_{sym}$ and the single extremum is a maximum. It follows that
${\cal E}_0$ is ever decreasing from its maximum value ${\cal
E}_0|_{sym}$: its minimum is reached at the boundary values for $\xi_1$,
which corresponds precisely to ${\cal E}_0|_{asym}$.

The last step consists in including the terms of order $\lambda_i^n \log
\lambda_i^2$ which were discarded until now. In the toy model considered
by Nambu, it is precisely these terms with $n=2$ which generate a
non-zero value for the Yukawa coupling which was zero in the previous
approximation. It is easy to check that, keeping these logarithmic terms
as in the second of eqs.(\ref{eq:effM4}), one generates terms of order
$\xi_i \log \xi_i$ , for $\xi_i \ll 1$, which are precisely the ones
needed to generate a large hierarchy among the two Yukawa couplings.

This toy model can easily be generated to the case of $N$ Yukawa
couplings. An interesting property of the Nambu mechanism is that in
this more general situation, one Yukawa coupling is naturally much
larger than all the remaining ones. In the simplest version of the
model, the latter Yukawas are equal but this degeneracy can easily be
lifted with intergenerational couplings.

\vskip 1cm
\section{Conclusion}
\vskip .5cm
We have studied how the mechanism proposed by Nambu to generate a
hierarchy between Yukawa couplings may be naturally implemented in
superstring low energy models.

Unlike the case of the standard model, it is natural in these models to
consider that Yukawa couplings are dynamical variables and the
corresponding minimisation amounts to a minimisation with respect to the
moduli of the underlying theory. Also, already at one loop,
the framework of softly broken supersymmetric theories makes it easier
to generate the right terms with the right signs. We stressed above that
the key ingredient for the success of the Nambu proposal in these models
is the cancellation of leading terms of order $m_{3/2}^2$ in $Str M^2$.
Then this supertrace is necessarily of order $m_{3/2}^4/M_{Pl}^2$. It
remains of course to be seen how such a constraint which is crucial for
the success of the mechanism may be dynamically generated.

In our work, we have refrained from applying these ideas to a realistic
situation. Our goal was to study how the Nambu proposal can be realized
in a string low energy model. This is also the reason why we have
considered solely the generation of a hierarchy among the Yukawa
couplings. However, we have seen above several times that this leads to
some other important issues in these models: supersymmetry breaking, the
generation of a hierarchy between $M_W$ and $M_{Pl}$ (or almost
equivalently $m_{3/2}$ and $M_{Pl}$), the problem of a cosmological
constant. For instance, our constraint (\ref{eq:quad}) most certainly
arises from a minimisation process that we have avoided to discuss here.
In the same line of thought, our potential depends on an overall scale
--~say $m_{3/2}$~-- which has to be determined dynamically. It must in
particular be checked that such a determination, in conjunction with our
minimisation procedure, does not lead to an instability of the theory.
The whole approach might for example be included into the ambitious
program undertaken by Kounnas, Pavel and Zwirner \cite{KPZ} who try to
determine dynamically $m_{3/2}$ and the top mass in a class of string
models where precisely $Str M^2$ cancels to leading order in
$m_{3/2}^2$.

Our own work is intended as a first look at the Nambu mechanism by
itself but it is clear that such issues as the ones listed above  will
now have to be addressed in order to make it a candidate for explaining
hierarchies in the observed spectrum. For example, in the realisation of
the Nambu idea, the key ingredient is the realisation of the quadratic
constraint among the Yukawa couplings. We have only shown here that such
a constraint is consistent with what is known of the string models which
are compatible with the low energy phenomenology. We certainly do not
feel that we
have a definite answer as to the origin of such a constraint. And much
progress remains to be done before a realistic scenario can be proposed.

\vskip 1cm
{\bf Acknowledgments.}
\vskip .5cm
We wish to thank T. Gherghetta, M. Peskin and F. Zwirner for valuable
discussions.

\vfill
\end{document}